# Unveiling spin-orbital angular momentum locking in photonic Dirac vortex cavities

Haitao Li,[1] Jiusi Yu,[1] Jiayu Fan,[1] Shijie Kang,[1] Bo Hou, [1] Zhi-Kang Lin,[2,5] Hanchuan Chen,[6] Jian-Hua Jiang,[2,3,4,*] and Xiaoxiao Wu[1,*]

1 Modern Matter Laboratory and Advanced Materials Thrust, The Hong Kong University of Science and Technology (Guangzhou), Nansha, Guangzhou 511400, Guangdong, China

2 School of Biomedical Engineering, Division of Life Sciences and Medicine, University of Science and Technology of China, Hefei, China

3 Suzhou Institute for Advanced Research, University of Science and Technology of China, Suzhou, Jiangsu, China

4 School of Physical Science and Technology, Soochow University, Suzhou, 215006, China

5 Department of Physics, The University of Hong Kong, Hong Kong, 999077, China

6 Key Lab of Advanced Transducers and Intelligent Control System, Ministry of Education and Shanxi Province, College of Physics and Optoelectronics, Taiyuan University of Technology, Taiyuan, 030024, China

* Correspondence author: xiaoxiaowu@hkust-gz.edu.cn (Xiaoxiao Wu) and jhjiang3@ustc.edu.cn (Jian-Hua Jiang).

**Abstract** Dirac vortices, originally studied in quantum field theories to predict localized zero-energy modes, were recently realized in photonics, leading to Dirac vortex cavities. With topological protection, Dirac vortex cavities offer robust single-mode large-area localized modes appealing for high-performance micro-lasers and other applications. As a spectrally-isolated single mode, the radiation of a Dirac vortex cavity mode was believed as having vanishing orbital angular momentum due to time-reversal symmetry. Here, we report the direct observation of orbital angular momentum radiation of a Dirac vortex cavity through spin-resolved measurements. Remarkably, we confirm the spin-orbital angular momentum locking in such radiation due to the spin-valley locking and inter-valley couplings. We demonstrate that the spin-orbital angular momentum locking is controlled by the chirality of the Kekulé modulation and propose design schemes for arbitrary-order single-mode OAM radiation.

## 1 Introduction

Topological cavities [1-3], such as Dirac vortex cavities, provide local modes with appealing properties such as single cavity mode with large area, spectral isolation, and resilience against fabrication imperfection. Recently, it was demonstrated that Dirac vortex cavities can support high-performance micro-lasers [4-10]. Alongside other topological cavity modes [11-18], Dirac vortex modes (DVMs) [19-25] which are the analogue of the Jackiw-Rossi zero modes [26-28], were realized in various photonic and phononic systems [29-31] as ideal cavity modes for various applications.

In our system, DVMs are spectrally isolated, nondegenerate bound states residing inside a band gap. For such an isolated mode in a reciprocal photonic system, radiative coupling to angular momentum channels is inherently balanced: reciprocity enforces equal coupling to counter-rotating angular momentum channels, so that any outgoing OAM flux is exactly canceled by its opposite counterpart. As a result, an isolated DVM is not expected to produce a net OAM in radiation, independent of the detailed field symmetry of the mode itself.

However, recent studies have reported OAM beams emission from isolated Dirac-vortex cavities [16], raising a fundamental question as to how a single, nondegenerate, and reciprocal topological bound state can give rise to net OAM radiation. This apparent contradiction constitutes the OAM paradox addressed in the present work.

Nevertheless, recent studies [26-28] suggest that DVMs can be regarded as the equal mixing of states with opposite spin or orbital angular momentum[29-32]. However, this intriguing nature and the underlying effects have not yet been confirmed directly in experiments. Moreover, we note that some earlier works on topological singularity mapping and pseudospin–orbit conversion [33] generate OAM through engineered geometric phases or real-space phase singularities. In contrast, the OAM demonstrated in this work originates intrinsically from a Dirac-vortex mode formed by a Kekulé mass vortex, where SOAM locking results from the interplay of intervalley coupling and spin–valley locking. Furthermore, recent advances in metalasers [34] derive structured emission through cavity anisotropy and non-Hermitian engineering, whereas our approach relies on a single protected topological mode and does not require gain–loss or multimode interference. These distinctions underscore the novelty and complementary nature of the present SOAM-locked emission mechanism.

In this work, the emergence of SOAM-locked OAM radiation in a Dirac vortex cavity follows a sequential valley-physics–driven mechanism. The induced modulation folds the K± valleys to Γ and introduces intervalley coupling, which hybridizes the opposite-valley components inside the DVM. Because each valley carries an intrinsic spin texture (spin–valley locking), the intervalley-hybridized DVM inherits well-defined SAM-dependent valley contributions. In photonic metasurfaces, each valley further possesses an intrinsic azimuthal phase winding, giving rise to valley–OAM locking. Consequently, the far-field radiation of a DVM exhibits SOAM locking: opposite SAM components radiate with opposite OAM signs. This mechanism explains how a nominally non-degenerate DVM can emit spin-resolved vortex beams despite its single-mode nature.

Previous approaches to structured-light or topological OAM generation—such as geometric-phase metasurfaces, pseudospin–orbit conversion, and real-space or momentum-space singularity engineering—rely on externally imposed phase gradients or multimode interference, and therefore do not reveal how a single, nondegenerate topological mode can emit OAM [33,35,36]. In contrast, Dirac-vortex cavities host a zero mode protected by a complex mass vortex [16,32], which has been widely assumed to exhibit no net OAM due to time-reversal symmetry and the absence of mode degeneracy [40].

Here we show that the Kekulé modulation and the associated intervalley coupling [37-39] fundamentally alter this picture: the interplay between valley–OAM locking [38] and spin–valley locking [41] enables a nondegenerate Dirac-vortex mode to intrinsically radiate SOAM-locked OAM, resolving the longstanding OAM paradox. Moreover, the radiated OAM order becomes tunable via a simple angular perturbation, without introducing additional modes. These features distinguish our work from all previous OAM and Dirac-vortex platforms and establish a new mechanism for generating structured light directly from a single topological mode.

## 2 Emergence of Spin-Orbital Locked DVMs

We start with a metasurface motherboard consisting of a dielectric slab and a triangular lattice of Y-shaped metallic patterns on the upper surface [Fig. 1(b)]. Calculation shows that the metasurface supports designer surface plasmons outside the light cone, i.e., forming Dirac cones at the $K_{\pm}$ points in the first Brillouin zone, as shown in Fig. 1(c). The Dirac points at 16.94 GHz (referred to as the Dirac frequency) are protected by the $C_{3v}$ symmetry of the metasurface [42,43]. The Dirac points will be gapped if such symmetry is broken, leading to a finite Dirac mass. When the unit cell is enlarged, as depicted by the yellow hexagon region in Fig. 1(b), the photonic bands are folded in the Brillouin zone which brings the two Dirac cones at the $K_{\pm}$ points to the Γ point, creating a double Dirac point with fourfold degeneracy (See SI Note S1).

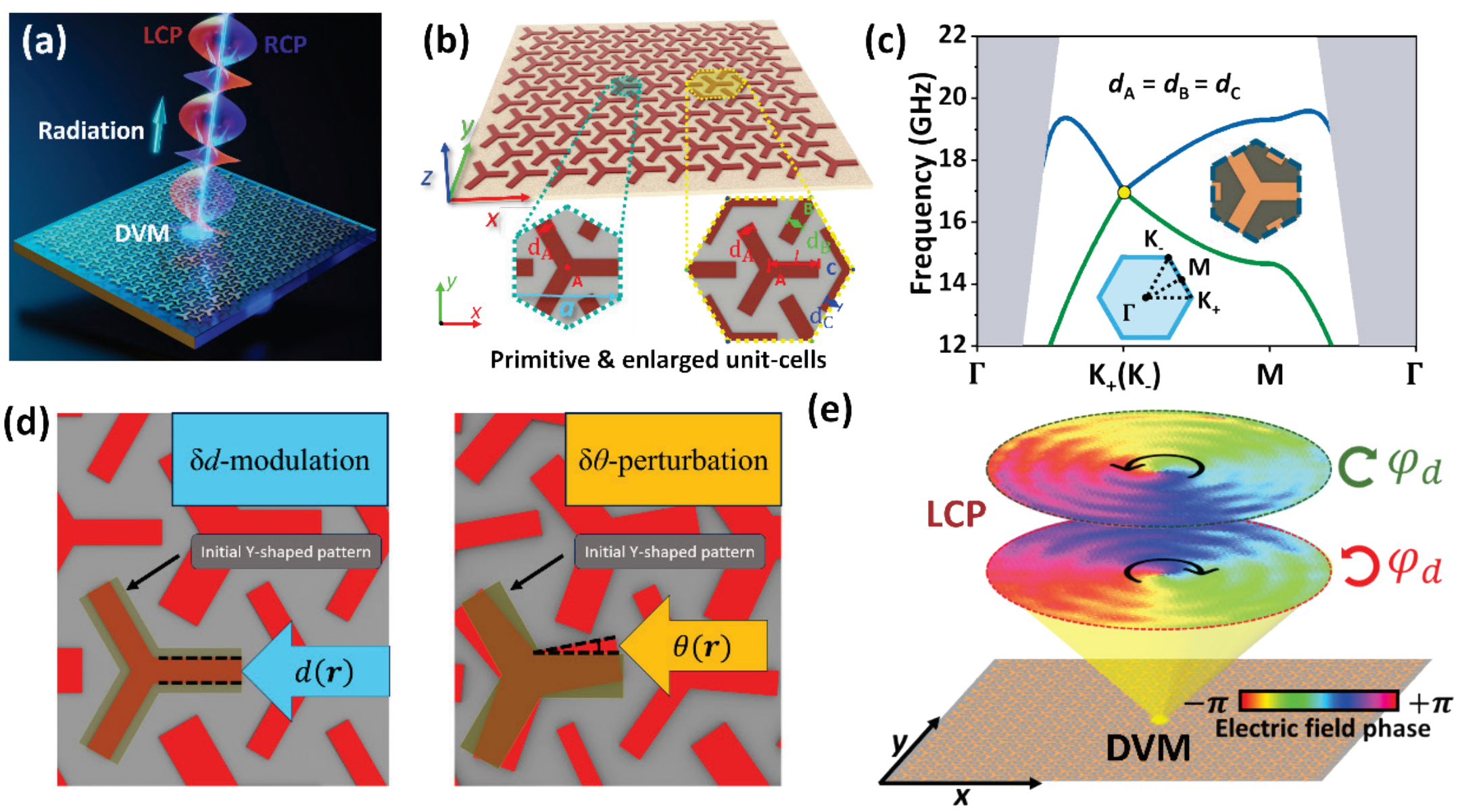

FIG 1. Metasurface platform for spin-locked OAM generation. (a) Schematic illustration of the main discovery: The DVM on a Kekulé metasurface radiates spin-locked OAM photons. (b) Initial metasurface without Kekulé modulation is a dielectric slab with thickness $h$=0.762 mm (relative permittivity $\varepsilon_r$=3.5) covered by a periodic metallic pattern (forming a triangular lattice with lattice constant $a$=5mm). Green-shaded region represents a primitive unit-cell. Yellow-shaded region is an enlarged unit-cell of which a zoomed-in view is given in inset where $l$=3 mm and $d_{\mathrm{A}}$=$d_{\mathrm{B}}$=$d_{\mathrm{C}}$= $d_0$=1 mm. (c) Photonic band structure calculated based on the initial primitive unit-cell. Yellow dot marks the Dirac points (at 16.94 GHz) at $K_{\pm}$ valleys. (d) Schematics of the $\delta d$-modulation and $\delta\theta$-perturbation applied to the Y-shaped metallic patterns on the metasurface. (e) Illustration of the reversal of the far-field OAM radiation for the LCP component with reversed winding of $\varphi_d$ in the Kekulé modulation.

The enlarged unit cell contains three inequivalent metallic Y-shaped patterns, labeled as A, B, and C. When their strip widths $d_{\mathrm{A}}$, $d_{\mathrm{B}}$, and $d_{\mathrm{C}}$ differ, the yellow hexagon cell becomes the primitive unit cell. This modulation induces a Dirac mass that gaps the double Dirac point (see SI Note S1). This Dirac mass is different from the one acquired by breaking the $C_{3v}$ symmetry. Engineering these Dirac masses is central to the design of Dirac vortex cavities. We find that Kekulé modulations of the Y-shaped patterns can lead to vortex-like patterns of the Dirac masses (i.e., Jackiw-Rossi-like Dirac vortices) which give rise to the DVMs.

A Kekulé pattern can be expressed as follows,

$$d(\boldsymbol{r}) = d_0 + \delta d \cos(\mathbf{K}\cdot\boldsymbol{r} + \varphi_d(\boldsymbol{r})), \#(1)$$

where $d_0$ is the strip width of the undeformed Y-shaped pattern, $\delta d$ and $\varphi_d$ parametrize the spatial dependent deformation. Here, the origin of the coordinate system is at the center of the Dirac vortex cavity. $\mathbf{K} = \mathbf{K}_{+} - \mathbf{K}_{-}$ is the so-called Kekulé vector.

To reveal the effect of the Kekulé modulation, we examine two cases with a constant $\varphi_d$ equaling to 0 or π, respectively, and omitting the spatial dependence. For these cases, the strip widths of the Y-shaped patterns, 'A', 'B', and 'C', are given by,

$$d_{\mathrm{A}} = d_0 + \delta d, \qquad d_{\mathrm{B}} = d_{\mathrm{C}} = d_0 - \frac{\delta d}{2}, \qquad \text{for } \varphi_d = 0,$$
$$d_{\mathrm{A}} = d_0 - \delta d, \qquad d_{\mathrm{B}} = d_{\mathrm{C}} = d_0 + \frac{\delta d}{2}, \qquad \text{for } \varphi_d = \pi. \#(2)$$

The double Dirac cone is gapped in both cases. However, the topological property of the band gap is distinct for these two cases as they correspond to distinct Dirac masses (see SI Note S1 for analysis). The effective Hamiltonian (see SI Note S2) around the Γ point can be written as

$$H(\boldsymbol{k}) = v_{\mathrm{D}}\tau_0 \otimes (k_x\sigma_1 + k_y\sigma_2) + (m_1\tau_1 - m_2\tau_2) \otimes \sigma_3, \#(3)$$

where $\boldsymbol{k} = (k_x, k_y)$ is the Bloch wave vector, $\sigma_i$ and $\tau_i$ represent two types of pseudospins described by the identity ($i = 0$) and Pauli ($i = 1,2,3$) matrices. The two types of pseudospins are connected to the sublattice pseudospin and the valley pseudospin in the honeycomb lattice model of the undeformed metasurface. $v_D$ is the Dirac velocity and $m_i$ ($i = 1,2$) stand for the Dirac masses induced by the intervalley couplings. It is noted that varying $\varphi_d$ from 0 to $2\pi$ results in the winding of the Dirac mass $m = m_1 + im_2$ in the complex plane, leading to a phase vortex of the Dirac mass which is often referred to the Jackiw-Rossi Dirac vortex (SI Note S14). In fact, a Dirac vortex cavity supporting $n$ DVMs can be achieved if $\varphi_d(\boldsymbol{r}) = n_d\alpha$ in Eq. (1) where $n_d$ is the order of $\delta d$-modulation and $\alpha = \arg(x + iy)$ is the polar angle of the coordinate vector $\boldsymbol{r}$. In this work, to achieve a single DVM, $n_d$ is set to ±1, and the sign of $n_d$ determines the winding direction of $\varphi_d$. [25]. We find that the winding direction of $\varphi_d$ determines the OAM direction of the far-field radiation at a given SAM [see Fig. 1(e)].

We now introduce an additional degree of freedom in the Kekulé modulation to engineer the DVM. As shown in Fig. 1(d), this $\delta\theta$ perturbation is achieved by rotating each Y-shaped pattern with a position-dependent angle

$$\theta(\boldsymbol{r}) = \delta\theta \cos[\mathbf{K} \cdot \boldsymbol{r} + \varphi_\theta(\boldsymbol{r})], \#(4)$$

where $\delta\theta$ denotes the strength of the perturbation and $\varphi_\theta(\boldsymbol{r}) = n_\theta\alpha$ describes the phase distribution. Notably, the integer $n_\theta$ represents the order of the $\delta\theta$ perturbation. For brevity, we refer to the modulation in Eq. (1) as $\delta d$ modulation and that in Eq. (4) as $\delta\theta$ perturbation. They are illustrated in Fig. 1(d). It is worth noting that the $\delta\theta$ perturbation plays a fundamentally different role from the $\delta d$ modulation. While the $\delta d$ modulation creates a Dirac mass vortex and is solely responsible for the existence of a DVM, the $\delta\theta$ perturbation does not modify the mass term and therefore cannot create or eliminate the bound state. Instead, $\delta\theta$ introduces a controlled $\boldsymbol{r}$-dependent rotation of each Y-shaped pattern, which imparts an additional valley-selective geometric phase.
In momentum space, this rotation acts as a valley-dependent phase winding $e^{in_\theta\phi}$. As a result, $\delta\theta$ does not influence the topological confinement of the DVM but modifies the radiative coupling strength of each valley component into the far field. Because the $K^+$ and $K^-$ valleys possess opposite intrinsic phase windings, the imposed $n_\theta$-order winding accumulates into an effective orbital angular momentum $L = \pm n_\theta$ in the emitted beam, with opposite signs locked to the SAM. This explains why $\delta\theta$ governs the OAM order of the radiation while leaving the DVM itself unaffected

We start with a metasurface with $\delta d$ modulation and $\varphi_d = \alpha$ and $\varphi_\theta = 0$, resulting in a single DVM (see SI Note S5). Full-wave simulations reveal a mid-gap mode with a high quality (Q) factor (>5000) at the Dirac frequency of 16.94 GHz (see SI Note S3). The electric field amplitude profile in Fig. 2(a) shows the DVM is tightly confined to the cavity. Fig. 2(b) gives the normalized current density distribution on the Y-shaped patterns at the metasurface center, indicating that the DVM is dominated by the out-of-plane magnetic dipole oscillation [44] (see SI Note S4 for details).

To further analyze the field distribution, we perform the 2D fast Fourier transformation (2D-FFT) on the $x$-component of near-field electric field ($E_x$) obtained from the simulation. Fig. 2(c) shows that the DVM is mainly composed of contributions from the two valleys $K_\pm$, suggesting valley mixing due to intervalley coupling induced by the Kekulé modulation (see SI Note S6). We then analyze the $E_x$-component of far-field radiation in simulations. The results in Fig. 2(d) show that, in contrast, the DVM is composed of contributions around the Γ-point. This finding is reasonable as the global Kekulé modulation fold the contributions at the $K_\pm$ valleys to the Γ-point, in a way similar to the band folding effect when the unit-cell is enlarged.

Interestingly, when apply Fourier transformation to the $E_x \pm iE_y$ near-field components, we observe dominant contribution from only one of the two valleys [Figs. 2(e)-(f)], revealing the spin-valley locking, i.e., the $K_+$ and $K_-$ valleys have opposite SAMs (see SI Note S6). This spin-valley locking and the OAM-valley locking reported before (e.g., [38]) together suggest the SOAM coupling. The intervalley coupling due to the Kekulé modulation then lead to the SOAM locking in the far-field radiation, highlighting a distinctive property of DVMs [41].

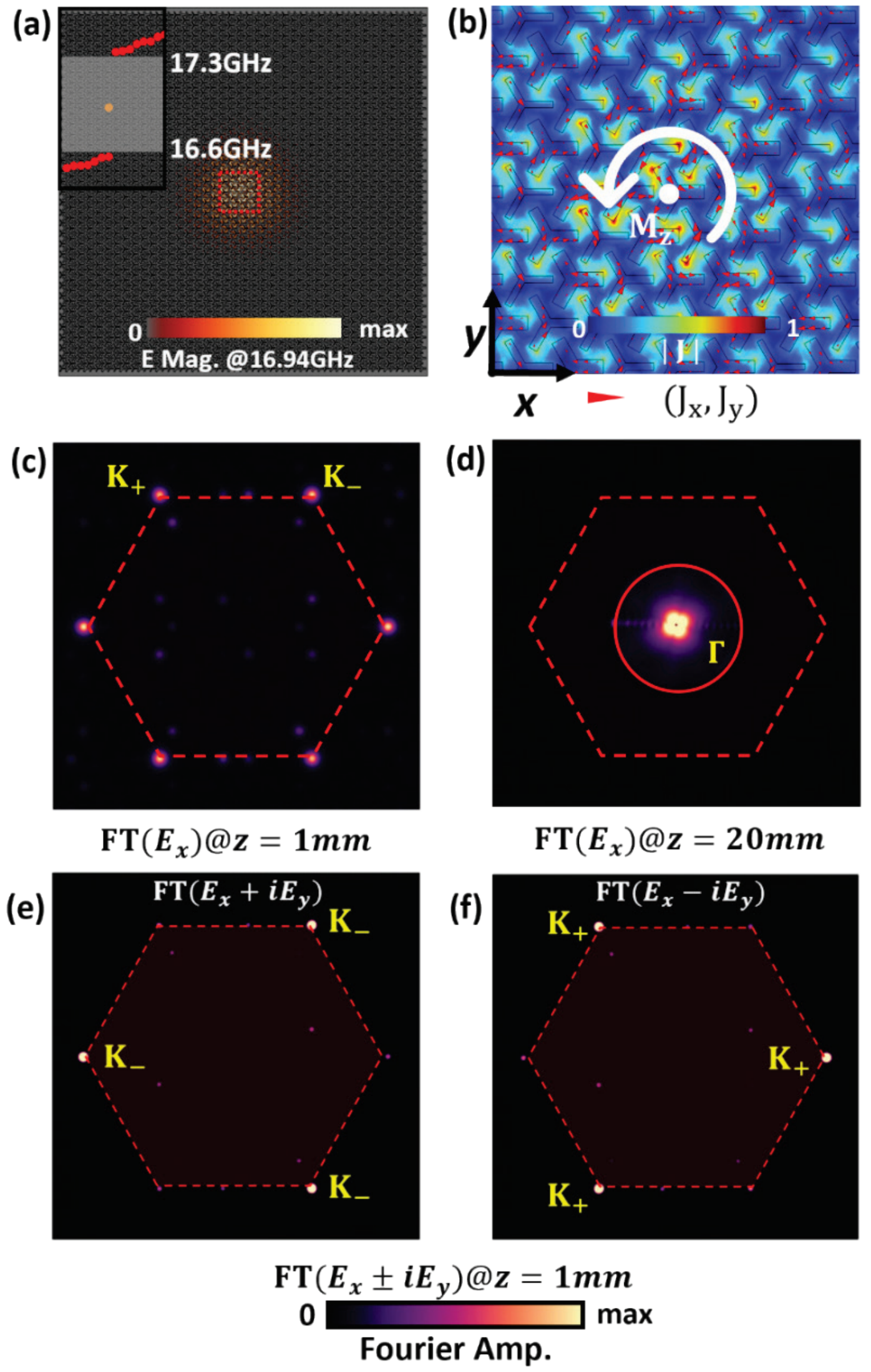

FIG 2. DVM induced by Kekulé $\delta d$ modulation on a metasurface with $n_d = +1$ and $\delta d = 0.4$ mm. (a) Electric field (magnitude |E|) profile of a DVM (at 16.94 GHz) at 1 mm above the metasurface. Inset: calculated spectrum of the metasurface where the DVM is labeled by orange dot. (b) Oscillating current distribution on the metasurface in the central region [dashed square in (a)]. The color map and arrows indicate magnitude and in-plane components of oscillating current $\mathbf{J} = (J_x, J_y)$, respectively. (c)-(d) The $k$-space Fourier spectra of the simulated near-field (c) and far-field (d) $E_x$-component at 16.94 GHz, as obtained via 2D Fourier transformation of the field profiles. (e)-(f) The $k$-space spectra for LCP (e) and RCP (f) components of the simulated near-field.

**3 Observation of the SOAM locking**

To probe the SOAM locking in the far-field radiation of DVMs, we detect the electric field at a plane 20 mm (60 mm) above the metasurface in simulations (experiments). We decompose the far-field radiation into the left- and right-circular polarized (LCP and RCP) components [45,46], corresponding to opposite SAMs of photons. As shown in Figs. 3(a) and 3(c), donut-shaped patterns are observed in the far-field electric field profile for both LCP and RCP components, indicating finite OAMs [47-50]. We design a particular setup to probe the radiation field [see SI Notes S4 and S5 for details]. The results in Figs. 3(b) and 3(d) show the measured radiation fields which exhibit donut-shaped profiles consistent with the simulation.

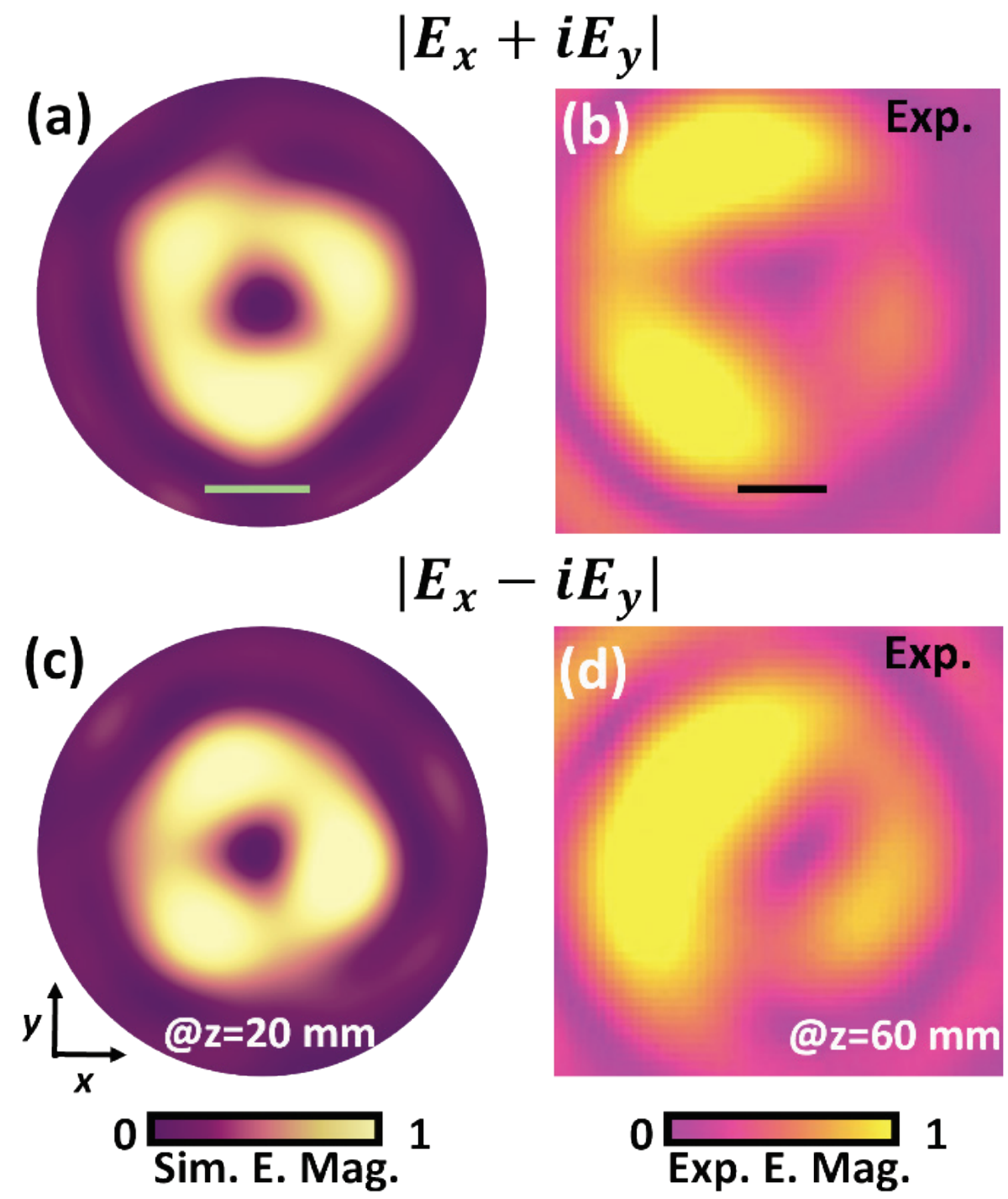

FIG 3. Experimental observation of vortex-beam far-field radiation from a DVM in FIG.2. (a) and (b) Electric field profiles (magnitude) of the far-field radiation obtained from simulations ($z$ = 20 mm) and experiments ($z$ = 60 mm), respectively, for the LCP component. The donut-shaped field profiles indicate the vortex beam radiation. (d)-(e) are similar to (b)-(c), but for the RCP component. The simulation and experimental results are obtained at frequencies 16.94 GHz and 16.87 GHz, respectively. The green and black scale bars stand for 20 mm.

We also find that if the $\delta\theta$-perturbation is introduced in this Dirac vortex cavity, the Q factor will be reduced. This effect is mainly because the $\delta\theta$-perturbation breaks the mirror symmetry of the original $\delta d$-modulation, making the DVM more radiative. Nevertheless, other properties such as the mode frequency is only slightly varied (<1.5%, see SI Notes S7 and S8). Interestingly, we find that the Q factor follows an exponential scaling relation with respect to $|\delta\theta|$. In our simulation, this relation is well fitted as (see SI Note S7)

$$Q = Q_0 \exp(-0.417|\delta\theta|), \#(5)$$

where $Q_0 = 12185$ is the Q factor of an ideal DVM without $\delta\theta$ perturbation.

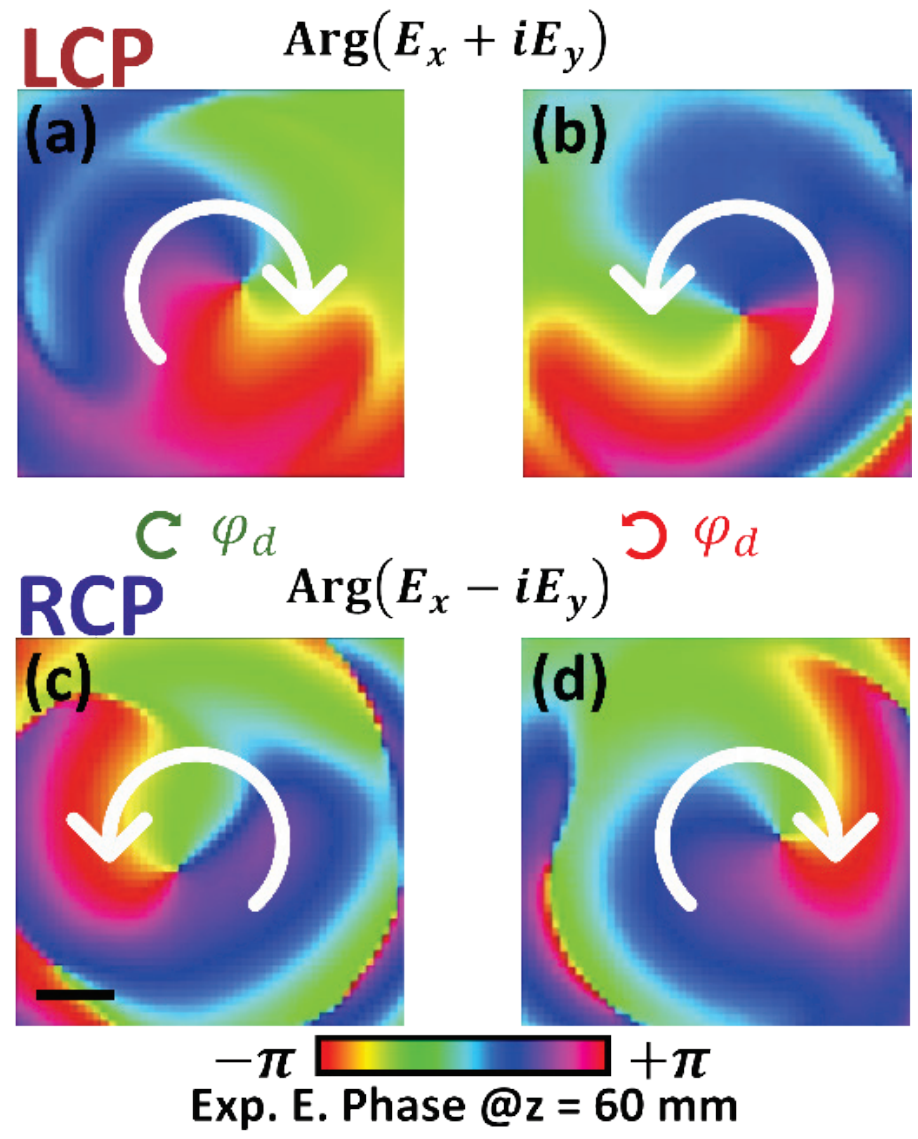

FIG 4. Experimental observation of spin-locked OAM radiation from the DVM with $n_d = \pm 1$ and $n_\theta = 1$ ($\delta d = 0.4$ mm, $\delta\theta = 10°$). (a)-(b) Experimentally measured phase distributions for the LCP component of the far-field radiation at 16.73 GHz on a plane 60 mm above the Kekulé metasurface with the charity of $\varphi_d$ being (a) clockwise ($n_d = -1$) and (b) counterclockwise ($n_d = +1$), respectively. (c)-(d) are similar to (a)-(b) but for the RCP component. Results indicate that the SOAM locking can be tuned by the winding direction of $\varphi_d$. See SI Note S12 for animations of the harmonic evolution of the phase vortices based on the experimental data. Black scale bar in (c) represents 20 mm.

We now directly confirm the SOAM locking and its tunability through the chirality of the Kekulé modulation (i.e., the winding direction of $\varphi_d$) in experiments. By decomposing the far-field radiation into LCP and RCP components, we observe clearly opposite phase vortices for opposite SAMs which are the smoking-gun signature of the SOAM locking. Furthermore, we uncover that for opposite $\varphi_d$ winding directions, the OAM for the same SAM reverses [see Figs. 4(a)-(b)]. Specifically, for the LCP component, clockwise winding of $\varphi_d$ results in OAM +1, whereas counterclockwise winding results in OAM −1. Figs. 4(c) and 4(d) show similar behavior for the RCP component, with reversed OAMs due to the SOAM locking in the DVM. In general, our results reveal that the topology of the radiation from the DVM undergoes a transition from a skyrmion to an anti-skyrmion configuration [51], controlled by the chirality of $\delta d$-modulation alongside $\delta\theta$ perturbation (see SI Note S15).

**4 Generation of Higher-Order OAM**

Interestingly, we find that the $\delta\theta$ perturbation can be engineered to yield higher-order OAM radiation from a single DVM which is appealing for on-chip OAM control. We present two examples with OAM L = 2 and 3 ~~(see SI Note S13)~~. The eigenfrequency spectra in Figs. 5(a) and 5(c) confirm that only a single DVM appears within the band gap (see SI Note S13 for more details). The far-field radiation patterns of the DVM, including amplitude and phase, are shown in Fig. 5(b), decomposed into the LCP (left column) and RCP (right column) components. The results indicate that the radiation carries OAM L = ±2 locked to opposite SAM components. The absolute value of L exactly corresponds to the winding number of the $\delta\theta$ perturbation. For the L = ±3 case, similar conclusions are found [Figs. 5(c)-5(d)]. This approach can be extended to arbitrary OAM orders, enabling the generation of vortex beams with on-demand OAM from a single cavity mode.

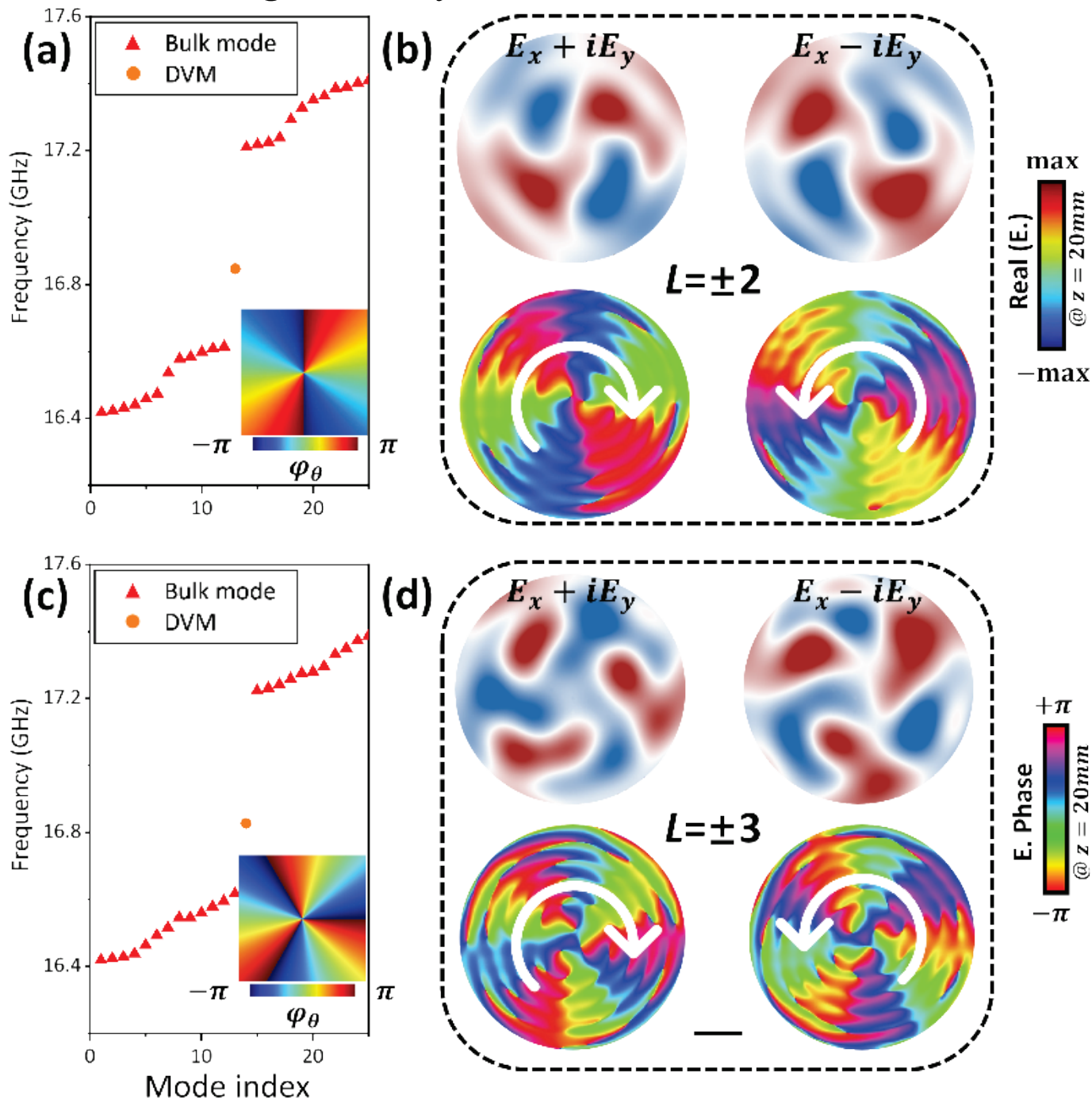

FIG 5. Higher-order OAM generation using a single DVM under $\delta\theta$ perturbations corresponding to cases where $n_\theta > 1$ (see SI Note S16 for more discussion). (a) and (c) Calculated eigenmode spectra of a metasurface under $n_\theta = 2$ (a) and $n_\theta = 3$ (c) $\delta\theta$ perturbation, respectively, where single DVM emerges at 16.85 GHz in the bandgap for both cases. (b) Real part (top row) and phase distribution (bottom row) on the far field cross section for DVM generated on 2nd $\delta\theta$ perturbation metasurface, for LCP (left column) and RCP (right column) components, respectively. The 2nd-order OAMs (L = ±2) can be clearly observed. (d) Similar to (c) but corresponding to the case of L = ±3. The black scale bar is 20 mm.

## 5 Conclusions and discussions

In summary, we establish that the emergence of SOAM-locked OAM radiation in a Dirac vortex cavity originates from a sequential mechanism governed by valley physics. The Kekulé δd modulation folds the originally inequivalent $K_+$ and $K_-$ valleys to the Γ point, enabling intervalley coupling and the hybridization of valley components within the DVM. Owing to the intrinsic spin–valley locking of each valley, the resulting hybridized DVM acquires well-defined, spin-dependent valley characteristics. Importantly, this valley–spin locking is dictated by the chirality of the Kekulé modulation phase, allowing deterministic switching of the valley–spin locking configuration by reversing the modulation chirality. Furthermore, in our metasurface platform, each valley carries an intrinsic azimuthal phase winding, leading to valley–OAM locking. As a consequence, the far-field radiation of the DVM exhibits robust spin–OAM locking, wherein opposite SAM components are bound to opposite OAM states. In addition, the $\delta\theta$ perturbation introduces higher-order degrees of freedom into the radiation of the DVM, enabling the emitted OAM order to be increased to $n_\theta$ without altering the single-mode nature of the cavity. This framework elucidates how a mid-gap isolated DVM can emit spin-resolved vortex beams despite its single-mode nature. Remarkably, the proposed Kekulé engineering provides a universal platform for on-demand OAM generation, scalable from microwaves to optical frequencies, and opens avenues toward on-chip topological OAM devices.

## 6 Acknowledgements

We acknowledge National Key R&D Program of China (No. 2022YFA1404400), National Natural Science Foundation of China (Nos. 12304348, 12074279, and 12125504), the Hundred Talents Program of Chinese Academy of Sciences, Gusu Leading Scientists Program of Suzhou City, Guangdong University Featured Innovation Program Project (No. 2024KTSCX036), Guangzhou Municipal Science and Technology Project (2024A04J4351), Guangdong Basic and Applied Basic Research Foundation (2025A1515011470), Guangdong Provincial Project (2023QN10X059), Guangzhou Higher Education Teaching Quality and Teaching Reform Engineering Project (2024YBJG087), and support from Wave Functional Metamaterial Research Facility of The Hong Kong University of Science and Technology (Guangzhou).